\documentclass[usenatbib,usedcolumn]{mn2e}
\usepackage{psfig,epsf}

\newcommand{\Msolar}{\mbox{${\; {\rm M_{\sun}}}$}}
\newcommand{\Lsolar}{\mbox{${\; {\rm L_{\sun}}}$}}
\newcommand{\MRC}{MRC~B1221$-$423}
\newcommand{\nii}{\mbox{[N$\,${\sc ii}]}}
\newcommand{\sii}{\mbox{[S$\,${\sc ii}]}}
\newcommand{\oii}{\mbox{[O$\,${\sc ii}]}}
\newcommand{\oiii}{\mbox{[O$\,${\sc iii}]}}
\newcommand{\oi}{\mbox{[O$\,${\sc i}]}}
\newcommand{\kms}{\mbox{${\;{\rm km\,s^{-1}}}$}}
\newbox\grsign \setbox\grsign=\hbox{$>$}
\newdimen\grdimen \grdimen=\ht\grsign
\newbox\laxbox \newbox\gaxbox
\setbox\gaxbox=\hbox{\raise.5ex\hbox{$>$}\llap
        {\lower.5ex\hbox{$\sim$}}}\ht1=\grdimen\dp1=0pt
\setbox\laxbox=\hbox{\raise.5ex\hbox{$<$}\llap
        {\lower.5ex\hbox{$\sim$}}}\ht2=\grdimen\dp2=0pt
\newcommand{\simlt}{\mathrel{\copy\laxbox}}

\newcommand{\rb}[1]{\raisebox{1.5ex}[0pt]{#1}}

\title[MRC B1221$-$423: a CSS source in a merging
  galaxy]{MRC~B1221$-$423: a compact steep-spectrum radio source in a
  merging galaxy\thanks{Based on observations made with ESO Telescopes
  at the La Silla Observatory, programs 70.A-0387(A) and
  66.A-0288(B)}}
\author[H. M. Johnston et al.]{Helen M. Johnston,$^1$\thanks{E-mail:
  H.Johnston@physics.usyd.edu.au}, Richard W. Hunstead$^1$,
  Garret Cotter$^{2,3}$ and 
  \newauthor
  Elaine M. Sadler$^1$\\  
  $^1$School of Physics, University of Sydney, NSW 2006, Australia \\
  $^2$Cavendish Laboratory, Madingley Road, Cambridge CB3 0HE \\
  $^3$Department of Astrophysics, University of Oxford, Keble Road,
  Oxford OX1 3RH}
\date{Received: } 

\begin{document}

\maketitle

\begin{abstract}
     We present {\it BVRIK}~images and spectroscopic observations of
     the $z=0.17$\ host galaxy of the compact steep-spectrum radio
     source MRC~B1221$-$423. This is a young ($\sim10^5$~yr) radio
     source with double lobes lying well within the visible galaxy.
     The host galaxy is undergoing tidal interaction with a nearby
     companion, with shells, tidal tails, and knotty star-forming
     regions all visible.  We analyse the images of the galaxy and its
     companion pixel-by-pixel, first using colour-magnitude diagrams,
     and then fitting stellar population models to the spectral energy
     distributions of each pixel. We also present medium-resolution
     spectroscopy of the system.
     
     The pixels separate cleanly in colour-magnitude diagrams, with
     pixels of different colours occupying distinct regions of the
     host galaxy and its companion.  Fitting stellar population models
     to these colours, we have estimated the age of each population.
     We find three distinct groups of ages: an old population ($\tau
     \sim 15$~Gyr) in the outskirts of the host galaxy; an
     intermediate-age population ($\tau \sim 300$~Myr) around the
     nucleus and tidal tail, and a young population ($\tau \simlt
     10$~Myr) in the nucleus and blue ``knots''.
     
     The spectrum of the nucleus shows numerous strong emission lines,
     including \oi\ $\lambda$6300, \oii\ $\lambda$3727, \sii\ 
     $\lambda\lambda$6716, 6731, H$\alpha$, and \nii\ 
     $\lambda\lambda$6548, 6583, characteristic of a LINER spectrum.
     The companion galaxy shows much narrower emission lines with very
     different line ratios, characteristic of a starburst galaxy.
     
     We have evidence for three distinct episodes of star formation in
     B1221$-$423. The correlation of age with position suggests the
     two most recent episodes were triggered by tidal interactions
     with the companion galaxy. The evidence points to the AGN in the
     centre of B1221$-$423 having been ``caught in the act'' of
     ignition.  However, none of the components we have identified is
     as young as the radio source, implying that the delay between the
     interaction and the triggering of the AGN is at least $3 \times
     10^8$~years.

\end{abstract}


\section{Introduction}
\label{sec:intro}

The onset of AGN activity in galaxies appears to be closely related to
starbursts, possibly triggered by mergers, which lead to large
increases in the amount of material being fed to the central black
hole, thereby triggering the radio emission.  Many radio galaxies show
signatures of tidal interactions, such as tails, bridges, shells, and
double nuclei \citep[see][for a review]{bh92}.  HST observations of
powerful radio galaxies at $z \sim 1$\ show disturbed, knotty
rest-frame UV emission in the host galaxies \citep*{lbr95,blr96}.

\begin{figure*}
     \centerline{\psfig{figure=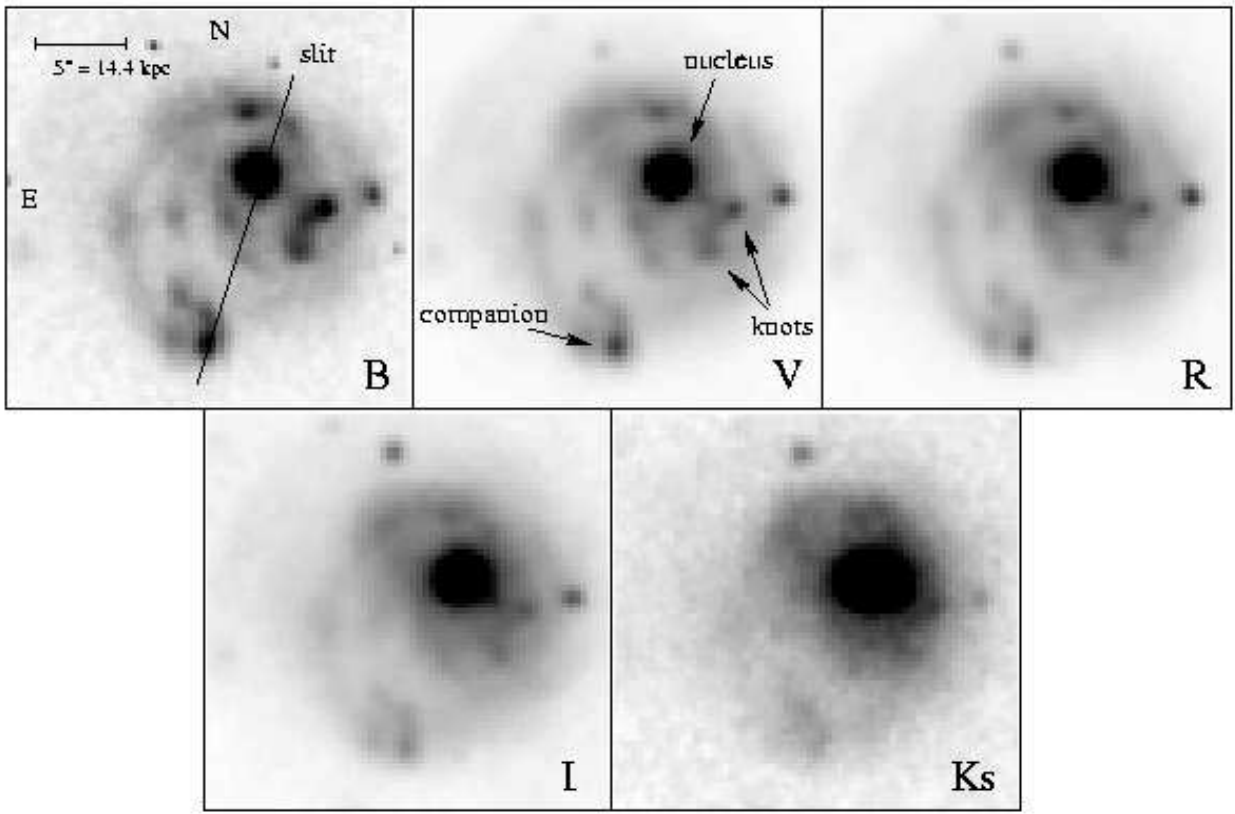,width=\textwidth,clip=t}}
     \caption{{\it BVRIK} images of MRC~B1221$-$423. Knots of
       star-formation and a tidal tail joining the galaxy to the
       companion are clearly visible. The size of each image is
       22~arcsec; the separation of the galaxy and the companion is
       10~arcsec ($1\,\mathrm{arcsec} = 2.88\,\mathrm{kpc}$\ assuming
       $H_0 = 71\;{\rm km\,s^{-1}\,Mpc^{-1}}$\ and
       $\Omega_\mathrm{M}=0.27$). The orientation of the slit for the
       spectroscopic observations (PA=$-$18\fdg5) is shown as the line
       joining the galaxy to the companion on the $B$\ image. The
       labels on the $V$\ band image indicate the various regions
       described in the text.}
     \label{fig:images}
\end{figure*}

Compact steep-spectrum (CSS) sources are bright but compact radio
galaxies, with sizes 1--20~kpc. Unlike the majority of powerful radio
galaxies, the radio source is wholly contained within the envelope of
the host galaxy. They have a steep spectral index, $\alpha < -0.5$\ 
($S_\nu \propto \nu^\alpha$), and a radio spectrum peaking below
500~MHz.  CSS sources are powerful ($P_{1.4} \ga 10^{25}\;{\rm
  W\;Hz}^{-1}$), and make up about 20\% of bright flux-limited
low-frequency radio catalogues like the 3CR catalogue \citep{ben62} or
the Molonglo Reference Catalogue \citep{lml+81}.  They typically have
redshifts between 0.1--2 \citep[see][for a review]{ode98}. The two
competing models for CSS sources are {\em (a)} that they are small in
size because the jets are ``frustrated'' by interaction with dense
surrounding gas; and {\em (b)} that they are young sources which will
eventually evolve into classical large double-lobed radio sources
\citep{ode98}.  Most evidence suggests that CSS sources are young
\citep[e.g.][]{ffs+90}, and will fade as they grow into large-scale
radio sources. Significant luminosity evolution is required to explain
the fact that the age of CSS sources is only 1\% that of the large
radio sources, yet 20\% or more of bright
centimetre-wavelength--selected samples are CSS sources
\citep[e.g.][]{sch74b,ka97,rtpw96}. The best evidence for the youth of
CSS sources is the detection of hotspot advance speeds of 0.1 to
0.3$c$\ in compact symmetric objects \citep[see][and
references therein]{con02}.

MRC B1221$-$423 is one of the nearest CSS sources, located in a host
galaxy with a redshift $z=0.1706$\ \citep{scrw93}. The radio source
has a steep spectrum ($\alpha = -0.85$) and double lobes with a
separation of only 1\farcs5, which lie well within the envelope of the
visible galaxy \citep*{shp03}.  The structure is suggestive of the
lobes seen in larger double sources; there is no evidence of a core.
\citet{shp03} estimated the age of the radio source to be $10^5$~yr,
using the tight correlation between the total kinetic power of the jet
and the narrow-line luminosity \citep{rs91}. At the distance of the
galaxy, 1~arcsec corresponds to a distance of 2.876~kpc, assuming $H_0
= 71\;{\rm km\,s^{-1}\,Mpc^{-1}}$\ and $\Omega_\mathrm{M}=0.27$.

The host galaxy is highly disturbed, showing signs of tidal
interaction with at least one close companion. An archival image taken
using the Anglo-Australian Telescope in 1\farcs4 seeing shows faint
concentric arcs to the north-east \citep{shp03}, evidence of past tidal
interaction.

We have obtained multi-colour images of MRC~B1221$-$423 in order to
study the change in properties of the stellar population across the
galaxy.  The use of spatially-resolved colours to study differences in
stellar populations across a galaxy has been shown to be an effective
way to study age distributions, localised extinction and star
formation history in galaxies. The technique was pioneered by
\citet{bot86}, who used a spatially resolved $(B-R)$ vs. $B$\ diagram
to examine the star formation history of NGC$\,$4449. More recently,
\citet{aef+99} used pixel colour-magnitude diagrams to investigate the
evolutionary history of galaxies in the \textit{Hubble Deep Field},
while \citet{kzc+00}, \citet{dlh+03} and \citet{kfp+03} used such
diagrams to study various nearby galaxies.

In this paper, we present and analyse five-colour images and optical
spectroscopic observations of MRC~B1221$-$423.  In \S\ref{sec:obs} we
describe the observations and their reduction.  In
\S\ref{sec:Modell-pixel-colo}, we analyse the images, first by means
of colour-magnitude diagrams, and then by fitting spectral energy
distributions to each pixel. In \S~\ref{sec:Surface-photometry} we
examine the surface-brightness profile of the galaxy.  In
\S\ref{sec:Spectroscopy-results} we discuss the spectroscopic results,
and then discuss the implications for the history of the galaxy in
\S\ref{sec:Conclusions}.

\section{Observations}
\label{sec:obs}

\subsection{Imaging}
\label{sec:imaging}

Four-colour optical images were taken on 2002 February 25 using SUSI2
on the New Technology Telescope (NTT) of the European Southern
Observatory with Bessell $B$, $V$, $R$ and $I$\ filters \citep{bes90}.
The spatial scale on the detector (two $2$k $\times 4$k EEV CCDs) is
0.161 arcsec/pixel. The conditions were photometric, and details of
the observations are given in Table~\ref{tab:obs}.  The equatorial
standard star field PG0942 from \citet{lan92} was observed to derive
the photometric transformations onto the Johnson-Cousins system.
Standard reductions were performed using the \textsc{iraf} software
suite.

An infrared image was obtained on 2002 December 23 through the
\emph{Ks} filter using the SofI instrument on the NTT. The field of
view using the Large Field objective was 4.9 arcmin, with a spatial
scale of 0.292 arcsec/pixel. The auto-jitter mode was used to obtain
fifteen separate images, with each image being an average of nine 7-s
images; thus the total exposure time on source was 945$\,$s.  The
telescope was moved by a random amount within a 100~arcsec box between
images.  Reduction of the images was done using the {\sc
  xdimsum}\footnote{{\sc DIMSUM} is the Deep Infrared Mosaicing
  Software package developed by Peter Eisenhardt, Mark Dickinson, Adam
  Stanford, and John Ward, and is available via ftp from {\tt
    ftp://iraf.noao.edu/iraf/extern/xdimsum/}} package in {\sc iraf}.
The data were flattened using a sky flat created by masking out
sources and combining the masked images; the final mosaic was then
created using {\tt xmosaic}.  The Persson standard star SJ 9154
\citep{pmk+98} was used to derive the photometric transformation.

Finally, all five images were registered onto the same pixel scale, by
fitting the positions of stars common to each image with linear
transformations, including terms for translation and scaling. The
$Ks$-band image was re-binned to the same scale, and a term for
rotation also included. The images were convolved with gaussians to
match the image size on the worst-seeing image (the $Ks$-band image,
with measured FWHM of 1.0 arcsec), and then binned $3\times3$, so the
pixel scale of the final images is 0.483 arcsec/pixel, which
critically samples the seeing disk. The registration of the images is
accurate to about 0.2 pixels, as measured from the positions of field
stars in the final images.

The final set of images is shown in Figure~\ref{fig:images}.  The
$B$-band samples rest-frame wavelengths in the range 3340--4180\AA, or
essentially rest-frame $U$-band.

\subsection{Spectroscopy}
\label{sec:spect}

\MRC\ was observed on four occasions, in 2001 April and 2003 August,
using the Double Beam Spectrograph on the ANU 2.3-m telescope at
Siding Spring Observatory. The detectors were two SITe $1752 \times
532$\ CCDs with 15-$\mu$m pixels. A complete log of observations is
given in Table~\ref{tab:obs}. For the observations in 2001, a dichroic
filter with a cross-over wavelength of 5500$\;$\AA\ was used to split
the light into the two arms of the spectrograph.  The observation in
2003 August was made with a low resolution grating and a plane mirror
in place of the dichroic, so that all the light was sent to the blue
arm of the spectrograph in order to obtain complete spectral coverage.
Pairs of 1200$\;$s or 1800$\;$s exposures were bracketed with CuAr
arc-lamp exposures. A slit of width 1.5 or 2~arcsec was used, oriented
at a position angle of $-$18\fdg5 so as to include light from the
companion on the slit. On 2001 April 16, the position angle was
$-$16\degr. The spatial scale on the detector was 0\farcs91/pixel,
corresponding to a linear scale of 2.6$\;$kpc per pixel.

\begin{table}
\caption{Journal of observations of MRC~B1221$-$423. The imaging
  observations were taken with the New Technology Telescope at the
  European Southern Observatory; the spectroscopic observations were
  taken using the Double Beam Spectrograph on the ANU 2.3m telescope
  at Siding Spring Observatory. The columns show the date of the
  observation, the wavelength range and the resolution in each of the
  spectrograph arms, and the exposure time. The seeing for the
  spectroscopic observations was 1\farcs2--1\farcs5.
  \label{tab:obs}}\addtolength{\tabcolsep}{-1pt}
\begin{tabular}{llcc.}
\hline
\multicolumn{5}{c}{\textit{Imaging observations}} \\
UT Date & Instrument & Filter & \multicolumn{1}{c}{$t_\mathrm{exp}$} & \multicolumn{1}{c}{Seeing} \\
        &            &        & \multicolumn{1}{c}{(s)}           & \multicolumn{1}{c}{($''$)} \\
\hline
2002 Feb 25 & SUSI2 & $B$ & 600 & 0.9 \\
            &       & $V$ & 600 & 0.7 \\
            &       & $R$ & 400 & 0.6 \\
            &       & $I$ & 900 & 0.8 \\
2002 Dec 23 & SofI  & $Ks$ & 945 & 1.0 \\
\\
\end{tabular}
\begin{tabular}{l r@{--}l r@{--}lccc r}
\hline
\multicolumn{8}{c}{\textit{Spectroscopic observations}} \\
UT Date     & \multicolumn{4}{c}{$\lambda$\ range (\AA)} &
\multicolumn{2}{c}{FWHM (\AA)} & \multicolumn{1}{c}{$t_{\mathrm
    exp}$} \\
& \multicolumn{2}{c}{blue} & \multicolumn{2}{c}{red} & blue & red & (s) \\
\hline
2001 Apr 16 & 3180 & 5800 & 5880 & 7510 & 4.6 & 2.3 & 7200 \\
2001 Apr 17 & 3180 & 5800 & 6610 & 8550 & 4.6 & 2.3 & 5400 \\
2001 Apr 18 & 3180 & 5800 & 6610 & 8550 & 4.6 & 2.3 & 5400 \\
2003 Aug 3  & 3180 & 10370 & \multicolumn{2}{c}{--} & 9.7 & -- & 2400 \\
\hline
\end{tabular}
\end{table}

\begin{figure*}
     \centerline{
       \psfig{figure=figure2a.ps,width=8cm,clip=t}
       \hfill
       \psfig{figure=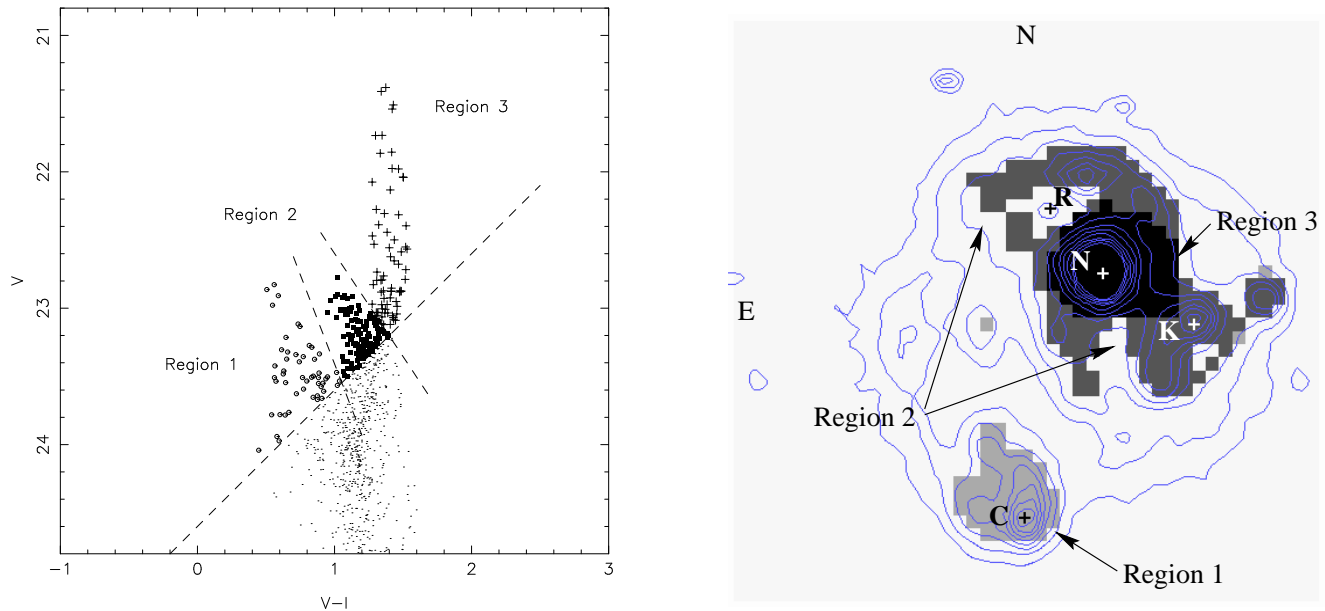,width=8cm,clip=t}
       }
     \caption{\emph{(a) }$V$\ vs. $V-I$\ colour-magnitude diagram for
       individual pixels in B1221$-$423. The units are magnitudes per
       pixel, where each pixel is 0.483 arcsec on a side.  All pixels
       brighter than $V=25$\ are shown as dots. Three distinct
       ``plumes'' of pixels are seen; these correspond to the three
       regions in \emph{(b)}.  The sky brightness was
       $V=22.8\,\mathrm{mag}\,\mathrm{pixel}^{-1}$. The dashed line
       shows the relation $(V-I) + V =24.6$~mag, which we took as the
       boundary delimiting the regions. \emph{(b)} Image showing the
       three regions identified in the colour-magnitude diagram,
       overlain with contours of the $B$-band image
       (Fig.~\protect{\ref{fig:images}}). Light grey pixels are from
       region 1, dark grey from region 2, and black from region 3. The
       letters refer to model spectra shown in
       Fig.~\protect{\ref{fig:sampleSED}}, representing pixels in the
       nucleus (N), companion (C), knot (K) and a red region to the
       north-east of the nucleus (R).}
     \label{fig:regions}
\end{figure*}

The {\sc iraf} software suite was used to remove the bias and
pixel-to-pixel gain variations from each frame. As we had multiple
consecutive observations of the same object, cosmic ray events were
removed using the technique described by \citet{cro95}, as implemented
in {\sc figaro}. The spectra were straightened in {\sc figaro} so that
the dispersion ran exactly along rows of the image, then a
two-dimensional wavelength fit was performed to the arc images by
fitting a third-order polynomial to the arc wavelengths as a
function of pixel number, for each row of the image. These wavelength
solutions were copied to the object images, interpolating between the
bracketing arc exposures, and the data were rebinned so the
wavelength-pixel relation was linear and uniform across the image. The
sky background was subtracted from each image, by fitting to the sky
on either side of the galaxy, chosen to be well outside the wings of
the galaxy profile. The spectra were corrected for atmospheric
extinction, the telluric absorption features were removed by comparing
with the spectrum of a smooth-spectrum standard taken at similar
airmass, and the spectra were flux-calibrated. The nights were not all
photometric, so the flux calibration can only be regarded as
approximate.

\section{Analysis}
\label{sec:analysis}

\subsection{Modelling pixel colours}
\label{sec:Modell-pixel-colo}

\subsubsection{Colour-magnitude diagrams}
\label{sec:Colo-magn-diagr}

Having created a uniform set of images, with the same scale,
orientation and seeing, we can now use this dataset to study the
variation in the stellar population across the face of the galaxy. 

The pixel colours were first corrected for Galactic extinction, using
the extinction map of \citet*{sfd98}, which indicates $A_V =
0.331$~mag at the position of MRC B1221$-$423. The other wavebands
were corrected using the reddening law of \citet{rl85}.  The $(V-I)$
vs. $V$\ colour-magnitude diagram for individual pixels of
MRC~B1221$-$423 is shown in Figure~\ref{fig:regions}\emph{(a)}. Three
distinct ``plumes'' of pixels are apparent in this diagram. We have
mapped these three features, bounded on the lower edge by pixels
satisfying the relation $(V-I) + V \le 24.6$~mag, onto the image.
This defines three distinct regions, shown in
Figure~\ref{fig:regions}\emph{(b)}. The features in the
colour-magnitude diagram are clearly separate in the image. Pixels in
the bluest region, Region~1, correspond to the interacting companion,
with a hint of a trail of pixels back up the tidal tail (light grey
pixels in Figure~\ref{fig:regions}\emph{(b)}). The reddest pixels,
Region~3, come from the centre of the galaxy (black pixels); while the
intermediate plume of pixels, Region~2, corresponds to regions at
larger radial distance from the centre of the galaxy, including the
``knots'' in the $B$-band image (dark grey pixels).

The lower boundary line in Figure~\ref{fig:regions}\emph{(a)} is
arbitrary, and was chosen to show clearly the spatial separation of
pixels of similar colours.  Other features, such as the tidal tail,
also occupy distinct regions of the colour-magnitude diagram, though
the separation becomes blurred as the pixel flux density drops
significantly below the sky level (22.8 mag per pixel in the $V$\ 
band).  The same separation is seen for other colour-magnitude
combinations, but is strongest for widely-separated bands.

Thus the colour-magnitude diagram separates pixels into distinct
groups, with different regions of the diagram occupying different
areas of the galaxy and its interacting companion. The most obvious
interpretation of this effect is that the different groups represent
different stellar populations; in this case, the three groups we
define here probably represent the products of three distinct episodes
of star formation.

\subsubsection{Fitting spectral energy distributions}
\label{sec:Fitt-spectr-energy}

A colour-magnitude diagram does not use all the information we have
available. With five colours, we can go further in analysing the
stellar population: by fitting spectral models to the colours of each
pixel in turn, we can determine the age and reddening of the stellar
population across the face of the galaxy.

We used a purpose-written program to model the spectral energy
distribution (SED) of pixels, loosely based on the \textsc{hyperz}
code of \citet*{bmp00}. The method is as follows. An input stellar
population model is chosen, and the redshift is fixed at the known
redshift of the galaxy, $z=0.1706$. A set of model spectral energy
distributions is constructed by folding the input spectrum through the
{\it BVRIK} response functions, resulting in a grid of model fluxes,
with varying age and reddening along the axes, to be compared to the
observed {\it BVRIK} fluxes for each pixel. The difference between the
sum of the model fluxes and the observed fluxes in each bandpass is
computed, and the parameters of the best fit (age and reddening)
recorded.

\begin{figure}
     \centerline{\psfig{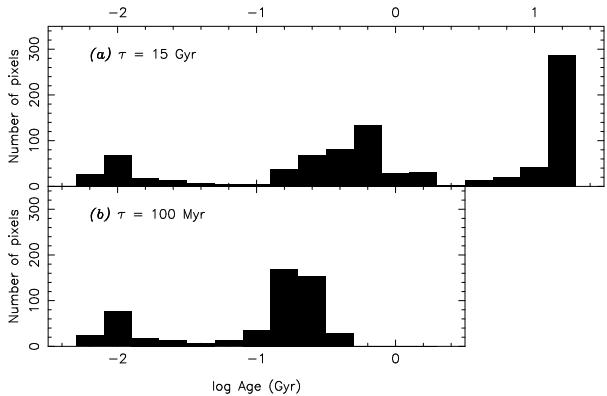}}
     \caption{Histogram of ages for each pixel, derived from the best-fit
       model. The top panel shows ages derived from a model with
       exponentially decaying star formation with a decay time of
       $\tau=15$~Gyr. The pixels are grouped into three distinct
       peaks, with ages of roughly 13~Gyr, 500~Myr and 10~Myr. The
       bottom panel shows a fit (excluding the pixels in the 15~Gyr
       peak) using a model with $\tau=100$~Myr. The two young
       populations are still present, with ages around 210~Myr and
       10~Myr. }
     \label{fig:age-hist}
\end{figure}

Motivated by the work of \citet{kfp+03} in modelling the stellar
populations in the merging galaxies known as the ``Antennae'', we
used as our initial model an old star-forming disk model with
exponentially decaying star formation on a timescale of $\tau=15$~Gyr.
The model was from the \textsc{gissel96} library \citep{bc93}, with a
Salpeter initial mass function with mass limits of 0.1 and 100\Msolar\ 
and solar metallicity, using the \citet{gs83} stellar spectral atlas.
A histogram of the fitted ages shows three distinct peaks, with ages
of 13~Gyr, 500~Myr, and 10~Myr (Figure~\ref{fig:age-hist}). We can
refine this by fitting a model with $\tau=100$~Myr to the ``young''
pixels, excluding those with ages above 5~Gyr
(Figure~\ref{fig:age-hist}, lower panel). The young pixels still show
two distinct age groupings, with ages of 210~Myr and 10~Myr and
younger.

\begin{figure}
     \centerline{\psfig{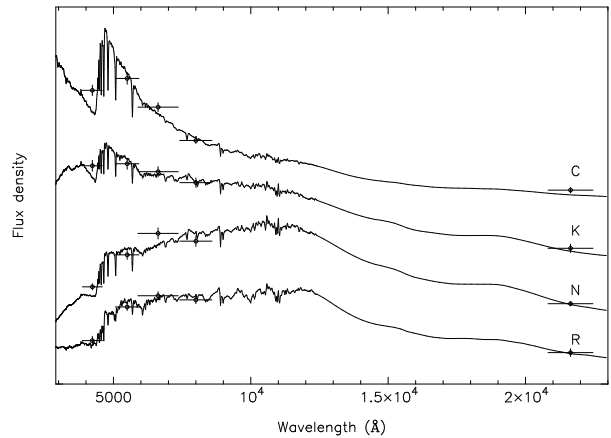}}
     \caption{Spectral models fitted to observed spectral energy
       distributions, for the four points denoted with a letter in
       Figure~\protect{\ref{fig:regions}}\emph{(b)}, and an
       exponentially-decaying star-formation model fitted, as
       described in the text. The squares represent the pixel fluxes
       to which the model SEDs are fitted. The horizontal bars
       represent the widths of the filters, the vertical bars the
       photometric errors on the pixel fluxes. The spectra are offset
       in the vertical direction for clarity. The pixels shown are
       labelled with letters in
       Figure~\protect{\ref{fig:regions}}\emph{(b)}; they represent
       pixels in the companion (C), one of the blue knots to the west
       of the nucleus (K), the nucleus (N), and a red region to the
       north-east of the nucleus (R). The parameters of the fit are
       shown in Table~\protect{\ref{tab:SEDfits}}.}
     \label{fig:sampleSED}
\end{figure}

Examples of the fitted spectral energy distributions are shown in
Figure~\ref{fig:sampleSED}. The observed spectral energy distributions
for the four points indicated with crosses in
Figure~\ref{fig:regions}\emph{(b)} are plotted, together with the
best-fit model spectra from the dual-age fit described above.  Point R
is in the ``old'' region, as indicated by the fit with the
$\tau=15$~Gyr model; the other three points were fitted using the
$\tau=100$~Myr model.  The fitted parameters are shown in
Table~\ref{tab:SEDfits}.

\begin{table}
\caption{Parameters of the fits to the spectral energy distributions
  for the four selected points shown in
  Figure~\protect{\ref{fig:regions}}\emph{(b)}: the fits are shown in
  Figure~\protect{\ref{fig:sampleSED}}. We fitted an
  exponentially-decaying star formation model with a timescale of
  either $\tau=15\,\mathrm{Gyr}$ or $\tau=100\,\mathrm{Myr}$, as shown
  in the second column. The best-fit age and reddening for each set of
  pixel fluxes is shown in columns 3 and 4.}
\label{tab:SEDfits}
\begin{tabular}{ll..}
\hline
Region & Model & \multicolumn{1}{c}{Age} & \multicolumn{1}{c}{$A_V$} \\
       &       & \multicolumn{1}{c}{(Myr)} & \multicolumn{1}{c}{(mag)} \\
\hline
\textbf{C}ompanion  & 100 Myr & 290 & 0.1 \\
\textbf{K}not       & 100 Myr &  28 & 1.5 \\
\textbf{N}ucleus    & 100 Myr & 128 & 2.0 \\
\textbf{R}ed region & 15 Gyr  & 15500 & 0.6 \\
\hline
\end{tabular}
\end{table}

\begin{figure}
     \centerline{
       \psfig{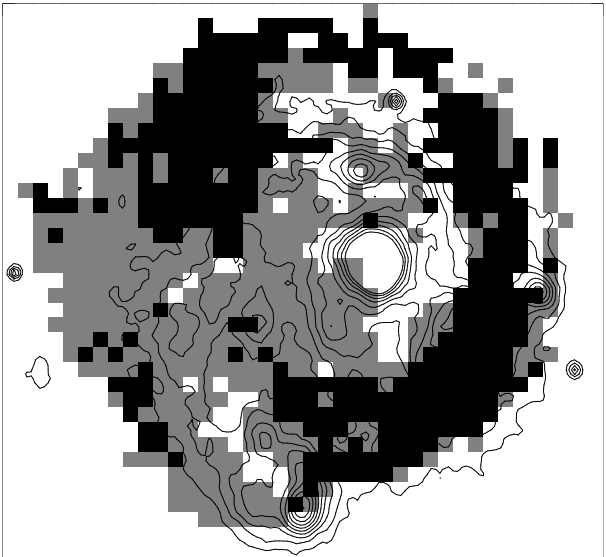}}
     \caption{Age image of MRC~B1221$-$423, created by fitting models
       with exponentially decaying star formation, with $\tau=15$~Gyr
       and $\tau=100$~Myr (see text for details). Pixels in the three
       peaks of the age histogram (Fig.~\protect{\ref{fig:age-hist}})
       are shown as different colours. The oldest pixels (ages $>
       5\,\mathrm{Gyr}$) are shown in black, pixels with ages between
       $80\,\mathrm{Myr}$\ and $5\,\mathrm{Gyr}$\ are shown as grey,
       and the young pixels (ages $< 80\,\mathrm{Myr}$) are shown as
       white.  Contours of the $B$-band image are overlain for
       comparison. The ages of pixels clearly correlate with
       morphological features of the galaxy. }
     \label{fig:age-image}
\end{figure}

We can examine the location of the populations of different ages by
plotting an age map of the galaxy. This is shown in
Figure~\ref{fig:age-image}. The ages of pixels clearly correlate with
morphological features of the galaxy (shown as contours for
comparison).  The old population is found through most of the galaxy,
and clearly represents the underlying stellar population of
MRC~B1221$-$423.  The intermediate age population, with ages around
500~Myr, is found in the companion and along the tidal tail joining
the galaxy to the companion, while the youngest population, with ages
$\sim 10$~Myr, is found in the centre of MRC~B1221$-$423 and in the
knots surrounding it, with a small group of young points in the
companion galaxy. In the $B$-band, the old population is contributing
30\% of the light, the intermediate-age population 46\%, and the young
population 24\%.

The fit is quite poor in the nuclear region (point N), especially for
the $V$\ and $R$\ bands. Examining the grid of models, there are two
solutions which are almost equally acceptable for points in this
region. The model shown in Figure~\ref{fig:sampleSED} has an age of
128~Myr and a reddening $A_V = 2.0$~mag ($\chi_\nu^2=1.7$); a second
solution, with a reduced $\chi^2$\ only marginally higher
($\chi_\nu^2=1.9$), has an age of 9~Myr and a reddening $A_V =
2.85$~mag. Since {\it BVRI} filters overlap and hence are not independent
of each other, the $\chi_\nu^2$\ values cannot be taken as a
quantitative measure of the goodness of fit; however, the relative
values indicate the second solution must be almost as good as the
first.

The split between these two solutions can be seen in
Figure~\ref{fig:age-image}, where some pixels in the nuclear region
are white (young, with ages $< 80$~Myr), while others are grey
(intermediate age, with ages between 80~Myr and 5~Gyr), despite having
similar colours (Figure~\ref{fig:regions}). This may well reflect a
real mix of ages in the nuclear region, so that a single age
population does not well describe the observed colours.  Alternately,
it could represent a spread in metallicity within the nuclear region,
since changes in both age and metallicity can lead to similar colours.

The spectrum of the galaxy shows many emission lines
(\S~\ref{sec:Spectroscopy-results}); however, as they are
comparatively weak (contributing at most 7\% of the flux), they do not
seriously distort our analysis.

\begin{figure}
     \centerline{
       \psfig{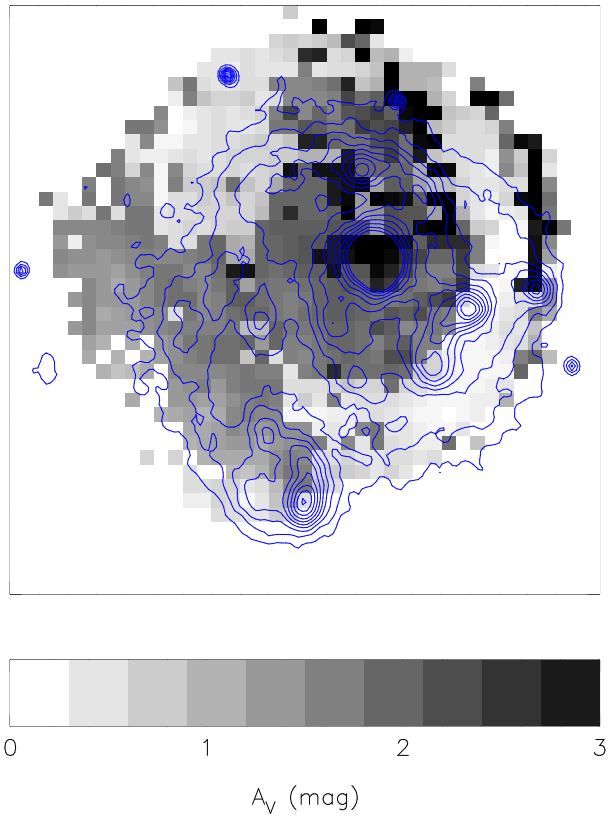}}
     \caption{Extinction image of MRC~B1221$-$423, from the same model
       fits as were used in Fig.~\protect{\ref{fig:age-image}}.  Black
       pixels correspond to heavily reddened regions, $A_V \ge 3$.
       Contours of the $B$-band image are overlain for comparison. }
     \label{fig:AV-image}
\end{figure}

We also derive the extinction across the galaxy from the same model
fits (Fig.~\ref{fig:AV-image}). Most of the region has $A_V < 1$, but
the centre of the galaxy and the blue regions to the west show
evidence for large amounts of dust, $A_V \sim 1$--2. This means that
the region with the youngest population -- the nucleus of the galaxy
-- requires the most internal reddening to fit the observed colours.

\subsection{Surface photometry}
\label{sec:Surface-photometry}

Since B1221$-$423 is a powerful radio source, we would expect its host
galaxy to be an elliptical galaxy; yet the optical image shows a
galaxy which looks like a spiral. Analysis of the light profile of the
galaxy seemed to be the best way to identify the shape of the
underlying stellar population.

We fit elliptical isophotes to the (unbinned) galaxy images, using the
task \texttt{ellipse} in \textsc{iraf}. We then attempted to fit
either de Vaucouleurs $r^{1/4}$-law profiles, appropriate for an
elliptical bulge, or a combined bulge plus exponential disk profile,
as expected for a spiral galaxy.

\begin{figure}
     \centerline{
       \psfig{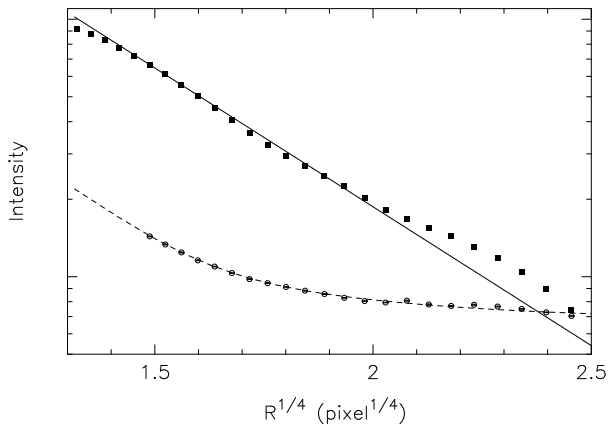}}
     \caption{$Ks$\ (filled squares) and $R$-band (open circles)
       luminosity profiles for MRC~B1221$-$423, plotted against
       $r^{1/4}$. The solid line shows the best-fitting de Vaucouleurs
       profile to the $Ks$-band data between 4 and 20 pixels ($1.4 <
       r^{1/4} < 2.1$). The dashed line shows a bulge+disk model fit
       to the $R$-band data, constraining the bulge to have the same
       radius as in the $Ks$-band, $R_e = 92$~pixels.
       }\label{fig:surfphot}
\end{figure}

The results of the analysis are shown in Figure~\ref{fig:surfphot}.
The $Ks$-band image is well described by an $r^{1/4}$-law profile, but
the $V$\ and $R$-band profiles are not. This implies that the host
galaxy does have a bulge, but it is being masked in bluer colours by
the younger populations.  We constrained the $V$\ and $R$-band
profiles to have a bulge of the same size as the $Ks$-band, with an
effective radius $R_e = 92$~pixels ($= 14.8$~arcsec). A de Vaucouleurs
profile with this radius plus an exponential disk is a good fit to
both the $R$\ and $V$-band light distributions (the $R$-band light
profile is shown in Fig.~\ref{fig:surfphot}). This bulge radius
corresponds to a half-light radius of 42~kpc. Correcting for the
extinction of $A_V=1.9$~mag as derived in the SED fitting, and
including $K$-corrections for the redshift $z=0.17$ from
\citet{pog97}, we can derive the intrinsic luminosity and colour of
the bulge. The bulge has $V-K = 3.1$, and a luminosity in the $V$-band
of $10^{12}\Lsolar$.

\subsection{Spectroscopy}
\label{sec:Spectroscopy-results}

\begin{figure*}
     \centerline{\psfig{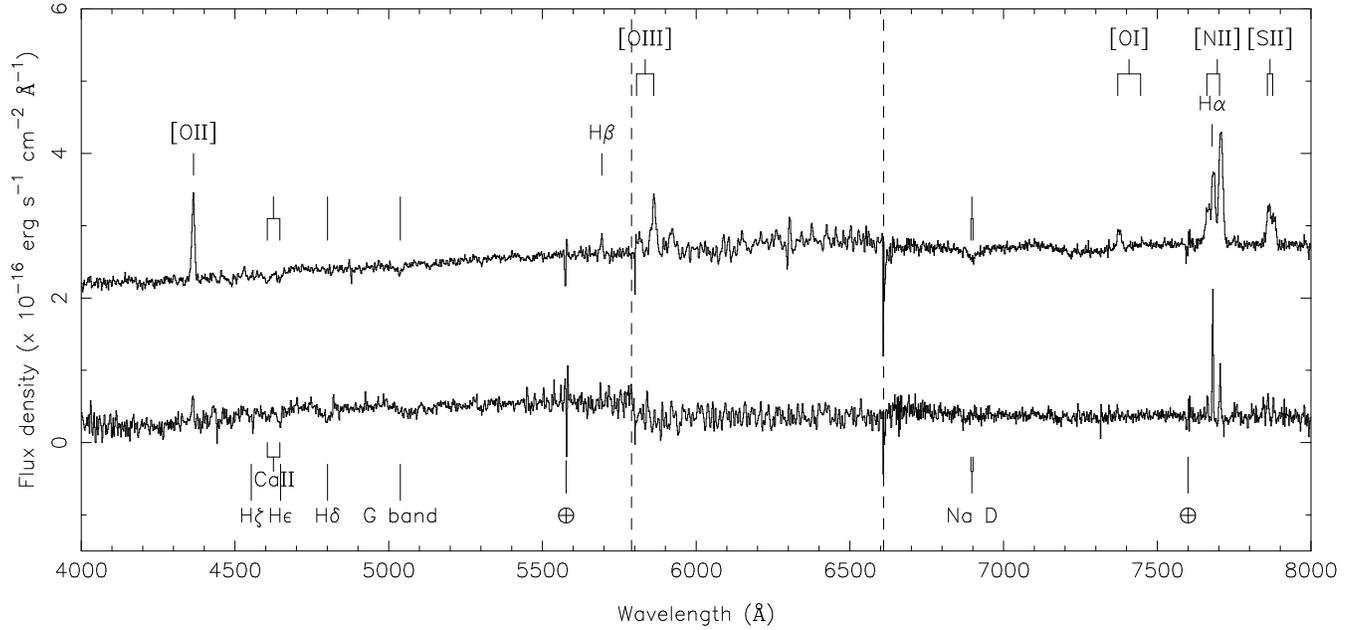}}
     \caption{Combined spectrum of the nucleus (top spectrum) and the
       companion (bottom spectrum), with emission and absorption lines
       identified. The nuclear spectrum has been offset in the
       \textit{y}-direction by $2\times 10^{-16}\,
       \mathrm{erg}\,\mathrm{s}^{-1}\,\mathrm{cm}^{-2}\,\mathrm{\AA}^{-1}$\ 
       for clarity. The region between the dashed lines is the lower
       resolution spectrum from 2003 August, which has much lower
       signal to noise; the data on either side are the high-quality
       summed spectra taken in 2001 April from the blue and red arms
       of the spectrograph.}
     \label{fig:spec}
\end{figure*}

We constructed a single spectrum (Fig.~\ref{fig:spec}) by summing the
spectral images from 2001, after aligning the spectra; we could only
use the data from 2001 April 17 and 18 for the red spectrum, since we
had a different wavelength setup for the first night
(Table~\ref{tab:obs}).

Emission is detected along the slit from the galaxy and companion.
The heliocentric redshift of the emission lines, determined from a
combined fit to the red side of the nuclear spectrum, is $z = 0.17068
\pm 0.00002$.  The lines show significant rotation, with an amplitude
of 90\kms\ along this axis.  The companion galaxy is found at a
velocity of $-105 \pm 20\kms$\ with respect to the nucleus.  The line
ratios can also clearly be seen to change along the slit, reversing
from the nucleus to the companion, with the nuclear spectrum showing
H$\alpha$\ weaker than \nii\ ($\log \nii/\mathrm{H}\alpha = +0.25$),
while in the companion \nii\ is weaker than H$\alpha$\ ($\log
\nii/\mathrm{H}\alpha = -0.35$).

We extracted the spectrum of the nuclear region by selecting the
central two rows, corresponding to a linear size of 5.2$\;$kpc. The
resulting spectra are shown in Figure~\ref{fig:spec}.  The line
properties were measured using the \texttt{specfit} package
implemented in \textsc{iraf} \citep{kri94}.  Gaussian profiles were
fitted to the lines, with a single velocity shift for all lines.  The
line fluxes and equivalent widths are given in Table~\ref{tab:lines}.

A spectrum was extracted for the companion galaxy, by summing together
seven rows at the position of the galaxy. This spectrum is also shown
in Figure~\ref{fig:spec}. This spectrum is clearly very different from
the spectrum of the nuclear region, with much narrower lines and very
different line ratios.

The nuclear spectrum shows the strong \oi\ $\lambda$6300, \oii\ 
$\lambda$3727 and \sii\ $\lambda\lambda$6716, 6731 emission lines
characteristic of the class of active galactic nuclei known as
``low-ionisation nuclear emission-line regions'', or LINERs
\citep{hec80}. We can further investigate the classification of the
different regions of the galaxy by using the classification diagrams
of \citet{vo87}.

From the ratio of the intensities of the H$\alpha$\ and H$\beta$\ 
lines, we can estimate the extinction towards the line-emitting
regions. For the nucleus of the galaxy, the observed ratio is 10,
compared with the expected ratio of 2.86 for a temperature
$T=10^4\,\mathrm{K}$, which implies an extinction $A_V = 3.7$~mag. For
the companion, the ratio H$\alpha$/H$\beta$ = 7.6, which implies an
extinction $A_V = 2.9$~mag.  These are both higher than the
extinctions we derived from modelling the spectral energy
distributions (Table~\ref{tab:SEDfits}), where we found $A_V =
1.9$~mag for the centre of the galaxy, and $A_V = 0.5$~mag for the
companion (plus 0.33~mag of local extinction).  However, since the
companion has both Balmer emission and absorption, the ratio of
H$\alpha$\ to H$\beta$\ may not be a reliable measure of the
extinction.

The nucleus of MRC~B1221$-$423 has the following line ratios,
corrected for the reddening as determined from the Balmer decrement:
\oi/H$\alpha$ = 0.3, \nii/H$\alpha$ = 1.9, \sii/H$\alpha$ = 1.0, and
\oiii/H$\beta$ = 6.5; these ratios match a LINER spectrum
\citep{vo87}. The ratio of the \oiii\ to H$\beta$\ flux is higher than
typical for a LINER spectrum, but since the \oiii\ flux is not well
determined (being seen only in the low-resolution spectrum from 2003
August) we give more weight to the high \nii/H$\alpha$\ ratio.

For the companion galaxy, the corrected line ratios are: \oi/H$\alpha$
= 0.07, \nii/H$\alpha$ = 0.5, \sii/H$\alpha$ = 0.4, and \oiii/H$\beta
< -0.05$\ (based on a marginal detection of H$\beta$\ and an upper
limit to \oiii), which match the spectrum of a starburst galaxy.

\begin{table*}
     \caption{Properties of detected emission lines and absorption
       lines in the spectra of the nucleus and companion galaxy
       (Fig.~\protect{\ref{fig:spec}}). The columns show the line with
       its rest wavelength, its equivalent width, integrated flux, and
       FWHM for the nuclear and companion spectra respectively.  Line
       properties were measured by fitting Gaussian profiles to the
       spectra; the values in the table are the properties of the
       fitted Gaussians.  The quantities with no errors (shown in
       italics) are linked to the corresponding quantities of another
       line, e.g. the flux ratio of the \oii~$\lambda\lambda$3727,
       3729 lines was fixed at 4.3, and their FWHMs constrained to be
       the same.  Negative values denote emission lines. The \oiii\ 
       $\lambda\lambda$ 5007, 4959 lines (marked with $^*$) were
       present only in the low-resolution spectrum of 2003 August.
       Lines which were not detected are indicated by a 3-$\sigma$\ 
       upper limit in the equivalent width column, calculated by
       $\mathrm{EW}_\mathrm{lim} = 3 \sqrt{2}\,
       \Delta\lambda/\mathrm{SNR}$, where $\Delta\lambda$\ is the
       dispersion (\AA/pix) and SNR is the signal-to-noise ratio of the
       spectrum at the location of the line; limits on doublets are
       shown as a single entry.}
     \label{tab:lines}
\begin{tabular}{l r@{$\;\pm\;$}l r@{$\;\pm\;$}l r@{$\;\pm\;$}l r@{$\;\pm\;$}l r@{$\;\pm\;$}l r@{$\;\pm\;$}l}
\hline
     & \multicolumn{6}{c}{\textbf{Nucleus}} &
     \multicolumn{6}{c}{\textbf{Companion}} \\
\\
     & \multicolumn{2}{c}{ }  & \multicolumn{2}{c}{Flux density}  &
     \multicolumn{2}{c}{ } &
     \multicolumn{2}{c}{ }  & \multicolumn{2}{c}{Flux density}  &
     \multicolumn{2}{c}{ } \\
Line & \multicolumn{2}{c}{EW} & \multicolumn{2}{c}{($\times 10^{-16}$} 
     & \multicolumn{2}{c}{FWHM} & 
     \multicolumn{2}{c}{EW} &
     \multicolumn{2}{c}{($\times 10^{-16}$} & \multicolumn{2}{c}{FWHM} \\
     & \multicolumn{2}{c}{(\AA)} &
    \multicolumn{2}{c}{\protect{$\mathrm{erg}\,\mathrm{s}^{-1}\,\mathrm{cm}^{-2}$)}} & 
     \multicolumn{2}{c}{(${\rm km\,s^{-1}}$)} &
     \multicolumn{2}{c}{(\AA)} & 
     \multicolumn{2}{c}{$\mathrm{erg}\,\mathrm{s}^{-1}\,\mathrm{cm}^{-2}$)} &
     \multicolumn{2}{c}{(${\rm km\,s^{-1}}$)} \\
\hline
\mbox{[O$\,${\sc ii}]} 3727 & $-40.9$ & 1.1 & $-7.3$ & 0.2 & 685 & 20 & 
   $-7.5$ & 1.4 & $-2.2$ & 0.4 & 500 & 130 \\
\mbox{[O$\,${\sc ii}]} 3729 & \multicolumn{2}{c}{$\mathit{-9.4}$} & \multicolumn{2}{c}{$\mathit{-1.7}$} & \multicolumn{2}{c}{$\mathit{685}$} & 
   \multicolumn{2}{c}{$\mathit{-1.8}$} & \multicolumn{2}{c}{$\mathit{-0.5}$} & \multicolumn{2}{c}{$\mathit{500}$} \\
H$\beta$  4861 &  $-3.1$ & 0.9 &  $-1.3$ & 0.4 & 360 & 120 & 
   $-1.9$ & 1.0 & $-1.0$ & 0.6 & 190 & 150 \\
\mbox{[O$\,${\sc iii}]} 4959$^*$ & \multicolumn{2}{c}{$\mathit{-4.3}$} & \multicolumn{2}{c}{$\mathit{-3}$} & \multicolumn{2}{c}{$\mathit{800}$} & \multicolumn{2}{c}{\rb{}} & \multicolumn{2}{c}{\rb{ }}& \multicolumn{2}{c}{\rb{ }}\\
\mbox{[O$\,${\sc iii}]} 5007$^*$ & $-12.9$ & 1.6 & $-9$ & 1 & 800 & 110 & \multicolumn{2}{c}{\rb{$< 6$}} & \multicolumn{2}{c}{\rb{-}} & \multicolumn{2}{c}{\rb{-}} \\
\mbox{[O$\,${\sc i}]}  6300 &  $-4.9$ & 0.5 & $-3.4$ & 0.4 & 560 & 60 & 
   \multicolumn{2}{c}{$< 0.7$} & \multicolumn{2}{c}{-} & \multicolumn{2}{c}{-} \\
\mbox{[N$\,${\sc ii}]} 6548 &  \multicolumn{2}{c}{$\mathit{-11.1}$} & \multicolumn{2}{c}{$\mathit{-8.5}$} & \multicolumn{2}{c}{$\mathit{580}$} & 
   \multicolumn{2}{c}{$\mathit{-3.3}$} & \multicolumn{2}{c}{$\mathit{-1.2}$} & \multicolumn{2}{c}{$\mathit{170}$} \\
H$\alpha$ 6563 & $-21.0$ & 0.9 & $-13.8$ & 0.4 & 480 & 20 & 
   $-18.0$ & 0.7 & $-7.6$ & 0.3 & 145 & 6 \\
\mbox{[N$\,${\sc ii}]} 6583 & $-33.2$ & 0.6 & $-25.3$ & 0.6 & 580 & 15 & 
   $-9.8$ & 0.6 & $-3.6$ & 0.3 & 170 & 15 \\
\mbox{[S$\,${\sc ii}]} 6716 &  $-11.2$ & 0.5 & $-8.3$ & 0.4 & 560 & 30 & 
   $-5.2$ & 0.9 & $-1.9$ & 0.3 & 200 & 30 \\
\mbox{[S$\,${\sc ii}]} 6730 &  $-8.0$ & 0.5 & $-5.9$ & 0.4 & \multicolumn{2}{c}{$\mathit{560}$} & 
   $-3.8$ & 0.7 & $-1.4$ & 0.3 & \multicolumn{2}{c}{$\mathit{200}$} \\
\\
Ca$\,${\sc ii} H 3933 & 5.8 & 0.7 & 1.3 & 0.1 & 1080 & 130 &
  3.5 & 1.9 & 1.4 & 0.8 & 790 & 420 \\
Ca$\,${\sc ii} K 3968 & \multicolumn{2}{c}{$\mathit{5.8}$} & \multicolumn{2}{c}{$\mathit{1.4}$} & \multicolumn{2}{c}{$\mathit{1080}$} &
  \multicolumn{2}{c}{$\mathit{3.5}$} & \multicolumn{2}{c}{$\mathit{1.9}$} & \multicolumn{2}{c}{$\mathit{790}$} \\
H$\zeta$         3889 & \multicolumn{2}{c}{$< 1.1$} & \multicolumn{2}{c}{-} & \multicolumn{2}{c}{-} & 3.2 & 1.8 & 1.3 & 0.7 & \multicolumn{2}{c}{$\mathit{860}$} \\
H$\epsilon$      3970 & \multicolumn{2}{c}{$< 0.9$} & \multicolumn{2}{c}{-}
     & \multicolumn{2}{c}{-} & \multicolumn{2}{c}{$< 1.5$} & \multicolumn{2}{c}{-} & \multicolumn{2}{c}{-} \\
H$\delta$        4101 & 1.2 & 0.8 & 0.3 & 0.2 & \multicolumn{2}{c}{$\mathit{1080}$} & 5.5 & 1.8 & 2.5 & 0.8 & 860 & 180 \\
G band           4304 & 4.1 & 0.7 & 1.2 & 0.2 & \multicolumn{2}{c}{$\mathit{1080}$} & \multicolumn{2}{c}{$< 0.7$} & \multicolumn{2}{c}{-} & \multicolumn{2}{c}{-} \\
Na D1 5889 & 1.5 & 0.3 & 1.0 & 0.2 & 620 & 180 & \multicolumn{2}{c}{ } & \multicolumn{2}{c}{ } & \multicolumn{2}{c}{ }  \\
Na D2 5895 & \multicolumn{2}{c}{$\mathit{0.8}$} & \multicolumn{2}{c}{$\mathit{0.5}$} & \multicolumn{2}{c}{$\mathit{620}$} & \multicolumn{2}{c}{\rb{$< 0.8$}} & \multicolumn{2}{c}{\rb{-}} & \multicolumn{2}{c}{\rb{-}}  \\
\hline
\end{tabular}
\end{table*}

The limited signal-to-noise ratio of our spectra preclude a more
detailed comparison with model population spectra, and do not allow us
to make direct comparisons with the models derived from the broad-band
images.  However, there are a few indicators of the underlying stellar
population which we can look at.

The $\lambda$4000-\AA\ break is a useful age-indicator for integrated
stellar populations. Older populations produce a discontinuity at
4000\AA\ due to the increase in stellar opacity produced by metal
lines shortward of 4000$\,$\AA\ in late-type stars.  We measured the
$\lambda$4000-\AA\ break index by comparing the mean flux density in
two windows above and below 4000$\;$\AA, taking the ratio of the mean
flux between 4050 and 4250$\;$\AA\ with the mean flux between 3750 and
3950$\;$\AA\ \citep{bru83}. For the nuclear spectrum, this index is
1.37. This small break index confirms the existence of recent star
formation in the nucleus of the galaxy \citep[e.g.][]{kwh+03b}.

Since the galaxy contains a powerful radio source, some fraction of
the excess blue light may arise from the AGN.  We detect the G~band in
the nuclear spectrum, with an equivalent width of 4.1\AA\ 
(Table~\ref{tab:lines}), which is consistent with the values seen in
normal elliptical galaxies \citep[e.g.][]{twf+98}; this means that we
are definitely detecting stars.  The detection of H$\delta$\ 
absorption, with an equivalent width of 1.2\AA\ means that at least
some of the blue excess is being contributed by intermediate-age
stars. We conclude that it is unlikely that the AGN contributes more
than about 50\% of the light.

\section{Conclusions}
\label{sec:Conclusions}

We have shown that the host galaxy of MRC~B1221$-$423 and its
interacting companion can be resolved into different stellar
populations. We have achieved this in two different ways, using
colour-magnitude diagrams, and by using models for the spectral energy
distribution of individual pixels. Both these methods separate the
pixels into distinct regions with different colours and/or ages. We
interpret this as indicating that the galaxy and its companion have
undergone several episodes of star formation, possibly associated with
the tidal interaction between the two.

The ages we deduce for the stellar population fall into three distinct
groups. An old population, with age $t \sim 15$~Gyr, occupies the
outskirts of the host galaxy. An intermediate age population, with
$t \sim 300$~Myr, is found near the nucleus and around the tidal
tail joining the two galaxies, while the youngest population, with
ages less than 10~Myr, is concentrated in the nucleus and the blue
regions to the west, as well as in the companion galaxy.

The location of these populations is not identical to the separation
suggested by the regions in the colour-magnitude diagram
(Figure~\ref{fig:regions}). This may be due to the effects of dust,
which our SED modelling suggests is concentrated in the nucleus of the
host galaxy. 

The relationship of the different stellar populations with the
underlying powerful radio source is much less clear. The age of the
source, estimated from the power of the radio jets and total stored
energy in the synchrotron plasma of the lobes \citep{rs91}, is only
$10^5$~yr \citep{shp03}.  This is much younger than the youngest
population we detected, with an age of about 10~Myr
(Section~\ref{sec:Fitt-spectr-energy}), and suggests there must be a
substantial time delay between the most recent burst of star formation
and the triggering of the radio source.

A possible scenario to explain the three component populations is as
follows: the tidal interaction began 300~Myr ago, which triggered star
formation in both the host galaxy and the companion. We see the
results of this episode of star formation as the intermediate-age
population.  Gas is driven down into the nucleus on timescales of
$\sim 10^8$~y, where it both cools to form stars, which we see as the
10~Myr-old population, and triggers activity in the central
supermassive black hole. Delays of this order are predicted in
theoretical calculations, \citep[e.g.][]{lpr88}, which give time
delays of a few$\times 10^8$~yr between the tidal interaction and the
gas reaching the centre of the galaxy.

This would be classed as a ``minor merger'', involving interaction of
the galaxy with a dwarf satellite. Such interactions have been shown
to be effective in feeding material into the nuclear regions
\citep{hm95}. This scenario for B1221$-$423 would fit with the idea
that low redshift radio galaxies are triggered by material falling
into a pre-existing supermassive black hole, in contrast with very
high redshift radio galaxies and quasars, which are often associated
with major mergers and the simultaneous formation of a massive bulge
and the central black hole.

\label{sec:Acknowledgments}

We thank Oliver Prouton for obtaining the optical images of
B1221$-$423, and the anonymous referee for helpful comments. RWH and
EMS acknowledge support for this project from the Australian Research
Council and the University of Sydney Sesqui grant scheme.


\end{document}